\documentclass{v16nufact}

\usepackage{graphicx}

\def\Journal#1#2#3#4{{#1} {\bf #2}, #3 (#4)}

\def\PRD{{\em Phys. Rev.} D}

\def\to{\rightarrow}

\def\be{\begin{equation}}
\def\ee{\end{equation}}
\def\bea{\begin{eqnarray}}
\def\eea{\end{eqnarray}}

\newcommand{\Photo}{}

\begin{document}
\vspace*{4cm}
\title{Status and future prospects for CLFV searches at BESIII}

\author{ Minggang Zhao (For the BESIII Collaboration) }

\address{School of Physics, Nankai University,\\
Tianjin, 300071, China}

 \maketitle
 \abstracts{
 Here we present the latest results of the charged Lepton Flavor Violation process 
 searches at the BESIII experiment in the decay of $J/\psi\to e\mu$, 
 using $(225.3\pm2.8)\times10^6$ $J/\psi$ events collected with the BESIII detector 
 at the BEPCII collider.
 An upper limit on the branching fraction of $\mathcal{B}(J/\psi\to e\mu)<1.6\times10^{−7}$ 
 (90\% C.L.) is obtained.
 The prospects and challenges with the future data are also discussed based on MC
 simulation.
 }

\section{Introduction}
As is well known, the Lepton Flavor Violation (LFV) is highly suppressed in the prediction 
of Standard Model (SM) by the finite but tiny neutrino masses. Its branching fraction is calculated 
to be at a negligible level and so far none has been found in all the historical experiments.
However, there are various theoretical models which can enhance the LFV effect large enough 
to be detected by the present experiments. Such as the SUSY grand unified theory \cite{ref::susy-gut}, 
SUSY with a right-handed neutrino \cite{ref::susy-right}, gauge-mediated SUSY breaking \cite{ref::susy-breaking},
SUSY with vector-like leptons \cite{ref::susy-vlepton}, SUSY with R-parity violation \cite{ref::susy-rparity}, 
models with Z$^\prime$\cite{ref::zprime}, or models with Lorentz non-invariance \cite{ref::noninvariance}.
Therefore, detection of such LFV decays could be taken as distinct evidence for new physics.

Experimentally, the search for LFV effect has been carried out in many ways, including lepton 
($\mu$,$\tau$) decays, pseudoscalar meson (K,$\pi$) decays, vector meson ($\phi$,$J/\psi$,$\Upsilon$) 
decays, etc. For example, a recent measurement based on $\mu^+\to\gamma e^+$ performed by the MEG 
Collaboration yields an upper limit of $\mathcal{B}(\mu^+\to\gamma e^+)<2.4\times10^{-12}$ \cite{ref::muon},
and a similar searching in $\tau$ decay by the BABAR Collaboration reports $\mathcal{B}(\tau^+\to\gamma e^+)<3.3\times10^{-8}$ \cite{ref::tau}.
Moreover, for neutral kaon and pion decays, the current results are $\mathcal{B}(K^0_L\to\mu^+ e^-)<4.7\times10^{-12}$ \cite{ref::kaon}
produced by the E871 Collaboration and $\mathcal{B}(\pi^0\to\mu^+ e^-)<3.8\times10^{-10}$ \cite{ref::pion} by the E865 Collaboration.
For LFV decays of vector mesons, despite having just collected relatively small data samples, 
evidences with better signal-significance have been observed, thanks to the simple background components.
The best measurement in $\phi$ decay, based on the data sample of 8.5 pb$^{-1}$ at the $e^+e^-$ 
annihilated energy region $\sqrt{s}=984-1060$ MeV, is obtained by the SND Collaboration in 2010 
setting upper limit of $\mathcal{B}(\phi\to\mu^+ e^-)<2.0\times10^{-6}$ \cite{ref::phi}.
In bottonium systems, based on about 20.8 million $\Upsilon(1S)$ events, 9.3 million $\Upsilon(sS)$ 
events, and 5.9 million $\Upsilon(3S)$ events accumulated with the CLEO-III detector, the CLEOIII 
Collaboration presented the most stringent LFV upper limit of 
$\mathcal{B}(\Upsilon(1S,2S,3S)\to\mu\tau)<\sim10^{-6}$ \cite{ref::cleo_measurement}.
In charmonium systems, the BESII Collaboration obtained $\mathcal{B}(J/\psi\to\mu e)<1.1\times10^{-6}$ 
\cite{ref::bes2-emu}, $\mathcal{B}(J/\psi\to e\tau)<8.3\times10^{-6}$ and $\mathcal{B}(J/\psi\to\mu\tau)<2.0\times10^{-6}$ 
\cite{ref::bes2-mutau} by analysing a data sample of 58 million $J/\psi$ events collected with the 
BESII detector, which are the best current upper limits on LFV effect in charmonium meson decays.
In this talk, we introduce the latest result from the BESIII Collaboration of searching for 
charged Lepton Flavor Violation decays based on baout 225 million 
$J/\psi$ events \cite{ref::jpsi_num_inc} collected at the BESIII detector.

\section{The Detector and Simulation}
\begin{figure}
\begin{minipage}{0.32\linewidth}
\centerline{\includegraphics[width=0.9\linewidth]{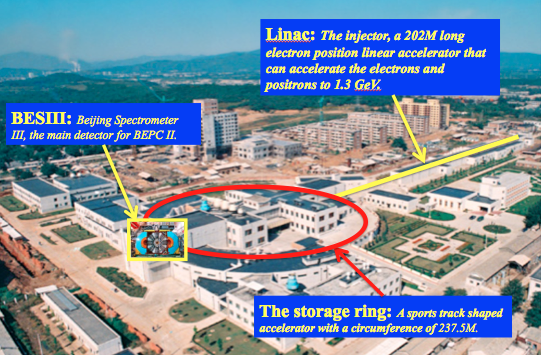}}
\end{minipage}
\hfill
\begin{minipage}{0.29\linewidth}
\centerline{\includegraphics[width=0.9\linewidth]{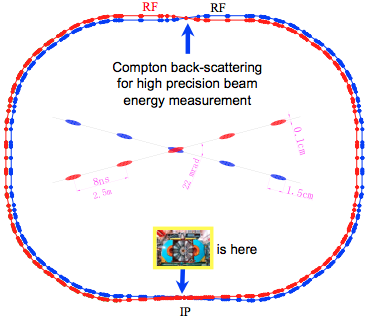}}
\end{minipage}
\hfill
\begin{minipage}{0.37\linewidth}
\centerline{\includegraphics[width=0.9\linewidth]{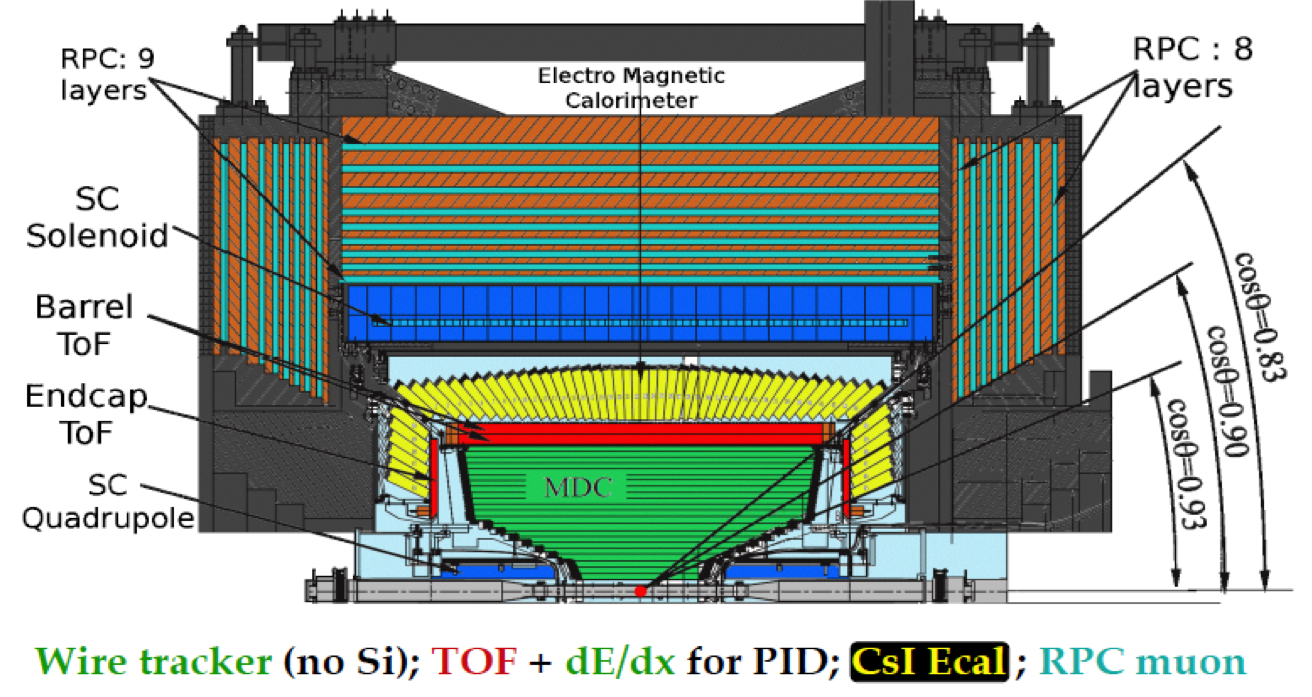}}
\end{minipage}
\caption[]{Illustration for the birdview of the experiment(left), the storage ring (center) and the detector (right)}
\label{fig::experiment}
\end{figure}

The BESIII experiment is composed of the LINAC, the BEPCII collider, and the BESIII detector 
\cite{ref::bes3_detector} (Fig. \ref{fig::experiment}), which is a large solid-angle magnetic 
spectrometer with a geometrical acceptance of 93\% of $4\pi$. It has four main components: (1) A 
small-cell, helium-based (40\% He, 60\% C$_3$H$_8$) main drift chamber (MDC) with 43 layers providing 
an average single-hit resolution of 135 $\mu$m, charged-particle momentum resolution in a 1.0 T magnetic 
field of 0.5\% at 1.0 GeV, and a ionization energy loss information ($dE/dx$) resolution better than 6\%. 
(2) A time-of-flight (TOF) system constructed of 5 cm thick plastic scintillators, with 176 detectors of 
2.4 m length in two layers in the barrel and 96 fan-shaped detectors in the end-caps. The barrel (end-cap) 
time resolution of 80 ps (110 ps) provides a $2\sigma$ $K/\pi$ separation for momenta up to $\sim1.0$ GeV. 
(3) An electromagnetic calorimeter (EMC) consisting of 6240 CsI(Tl) crystals in a cylindrical structure 
(barrel) and two end-caps. The energy resolution at 1.0 GeV is 2.5\% (5\%) and the position resolution is 
6 mm (9 mm) in the barrel (end-caps). (4) The muon system (MUC) consists of 1000 m$^2$ of Resistive Plate 
Chambers (RPCs) in nine barrel and eight end-cap layers and provides 2 cm position resolution.

The event selection and the estimation of backgrounds are optimized through Monte Carlo (MC) simulation. 
The {\sc geant4}-based simulation software BOOST \cite{ref::boost} includes the geometric description and 
material composition of the BESIII detector and the detector response and digitization models, as well
as the tracking of the detector running conditions and performance. The generic simulated events are generated 
by $e^+e^-$ annihilation into a $J/\psi$ meson using the generator {\sc kkmc} \cite{ref::kkmc} at energies 
around the center-of-mass energy $\sqrt{s}=3.097$ GeV. The beam energy and its energy spread are set according 
to the measurement of the BEPCII, and the initial state radiation (ISR) is implemented in the $J/\psi$ generation.
The decays of the $J/\psi$ resonance are generated by {\sc evtgen} \cite{ref::evtgen} for the known modes 
with branching fractions according to the world's average values \cite{ref::pdg2012}, and by {\sc lundcharm} 
\cite{ref::lundcharm} for the remaining unknown decays.

\section{Result of $J/\psi\to e\mu$}

At the BESIII experiment, the signal events are produced as $e^+e^-\to J/\psi$ at $\sqrt{s}=3.097$ GeV, and then
$J/\psi\to e\mu$, where the signal tracks are back-to-back opposite charged tracks with no extra EMC showers. 
The details of the event selection can be found in Ref.{ref::bes3-emu}. Based on a full simualtion to the physics
around the $J/\psi$ resonance, we found most of the backgrounds are from $J/\psi\to e^+e^-$, $J/\psi\to\mu^+\mu^-$, 
$J/\psi\to\pi^+\pi^-$, $J/\psi\to K^+K^-$, $e^+e^-\to(\gamma)e^+e^-$ and $e^+e^-\to(\gamma)\mu^+\mu^-$, in which 
one or more tracks are misidentified as muon or electron. To suppress these contamination events, several powerful 
criteria are employed.

\begin{figure}
\begin{minipage}{0.5\linewidth}
\centerline{\includegraphics[width=0.9\linewidth]{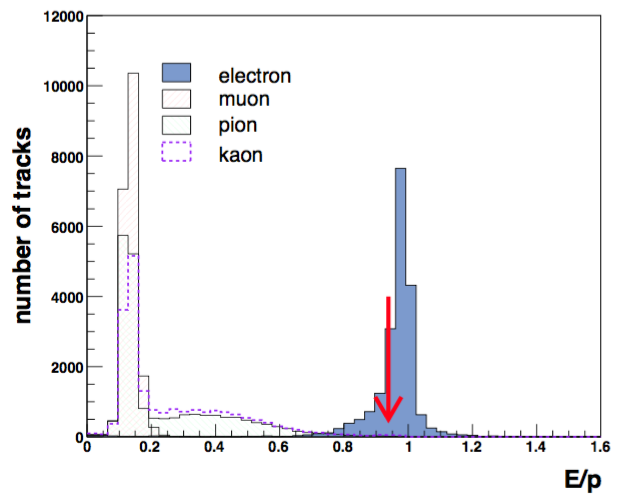}}
\end{minipage}
\hfill
\begin{minipage}{0.5\linewidth}
\centerline{\includegraphics[width=0.9\linewidth]{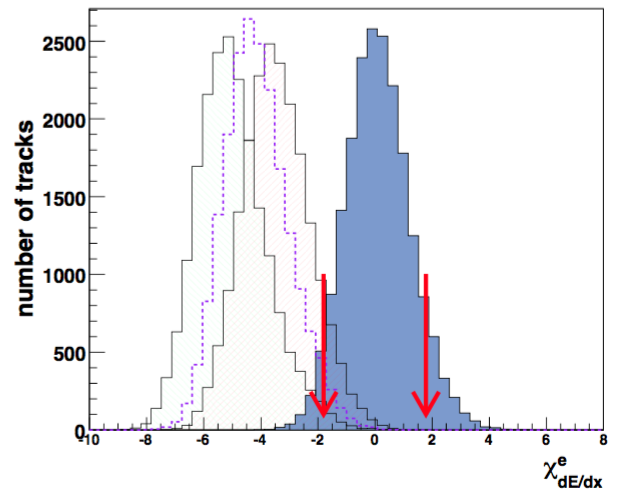}}
\end{minipage}
\caption[]{The distributions of $E/p$ (left) and $\chi^e_{dE/dx}$ (right) for the simulated electron, muon, pion and kaon samples.}
\label{fig::ep-dedx}
\end{figure}

For $e^{+}/e^{-}$ identification, there must be no associated hit in the MUC and the value of $E/p$ is required to 
be greater than 0.95 and less than 1.5, where $E$ is the energy deposited in the EMC and $p$ is the momentum measured 
by the MDC. The absolute value of $\chi_{dE/dx}$ from the $dE/dx$ measurement with electron hypothesis should be less 
than 1.8. Electron, muon, pion and kaon samples are simulated with MC method to investigate the above cut values, 
which are shown in figure \ref{fig::ep-dedx}, where the $E/p$ and $|\chi_{dE/dx}^{e}|$ distributions of electron can 
be well discriminated from other particles.

\begin{figure}
\begin{minipage}{0.5\linewidth}
\centerline{\includegraphics[width=0.9\linewidth]{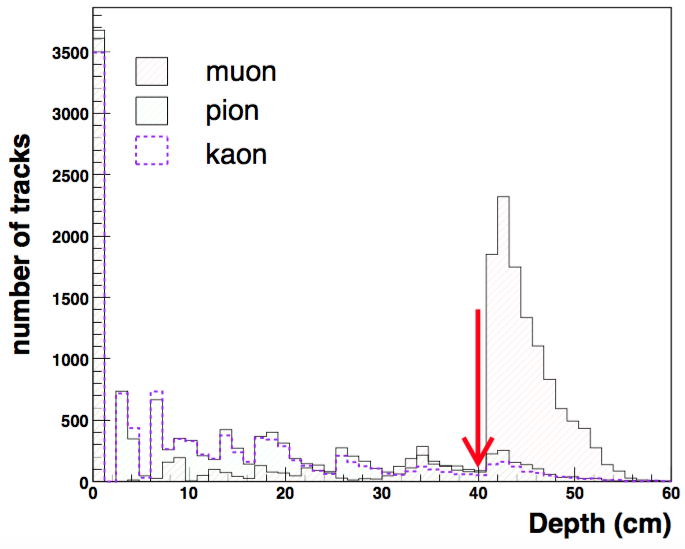}}
\end{minipage}
\hfill
\begin{minipage}{0.5\linewidth}
\centerline{\includegraphics[width=0.9\linewidth]{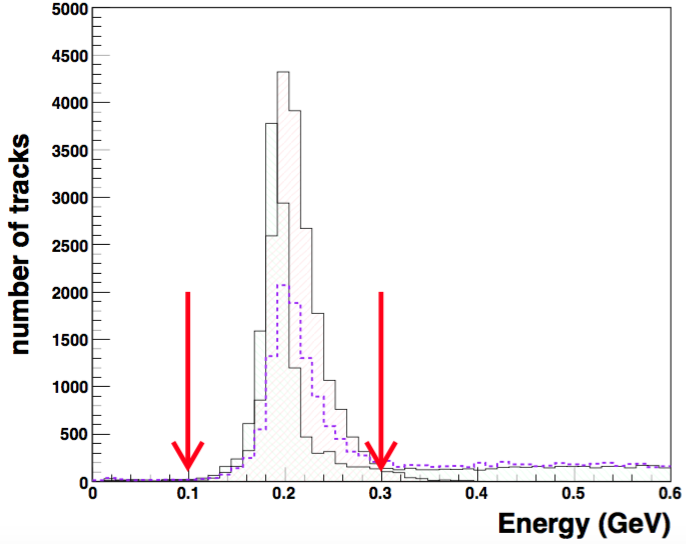}}
\end{minipage}
\caption[]{The distributions of the penetration depth in the MUC (left) and the deposited energy in the EMC (right) for 
    the simulated muon, pion and kaon samples.}
\label{fig::muonID}
\end{figure}

For $\mu+/\mu^-$ identification, the charged tracks in the active area of the barrel MUC $(|\cos\theta| < 0.75)$ are 
required to have a $E/p$ value less than 0.5, and the deposited energy in the EMC between 0.1 GeV and 0.3 GeV. In order 
to remove those tracks which are poorly reconstructed in the MUC, we require the penetration depth in the MUC larger 
than 40 cm and $\chi^2$ of track fitting in the MUC should be less than 100 if the track penetrates more than 3 detecting 
layers in the MUC. Furthermore, the $\chi_{dE/dx}$ value from the $dE/dx$ measurement with electron hypothesis must be 
less than -1.8. With the above simulated samples, distributions of the deposited energy in the EMC and the penetration 
depth in the MUC are shown in figure \ref{fig::muonID}, we can suppress the misidentification from pion and kaon with 
this two information.

After the above analysis, surviving events of $J/\psi \to e^+\mu^-$ are examined with two variables, $|\Sigma\bar{p}|/\sqrt{s}$ 
and $E_{vis}/\sqrt{s}$, where $|\Sigma\bar{p}|$ is the vector sum of the total momentum in one event, $E_{vis}$ is the total 
reconstructed energy, and $\sqrt{s}$ is the center-of-mass (c.m.) energy. A candiate event should be located in the signal 
box defined by 0.93 $\leq$ $E_{vis}/\sqrt{s}$ $\leq$ 1.10 and $|\Sigma\bar{p}|/\sqrt{s}$ $\leq$ 0.10, which corresponds to 
about 2 standard deviations of the variables determined by MC simulation.

\begin{figure}[htbp]
\begin{center}
\includegraphics[height=8cm,width=10cm]{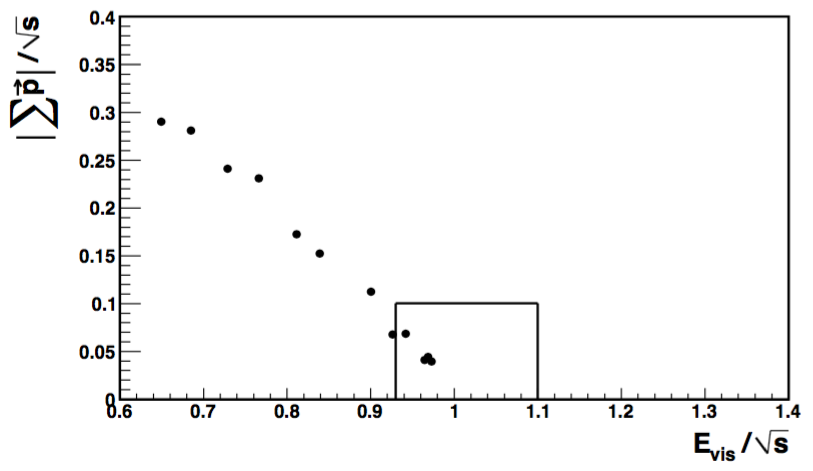}
\put(-170,-5){\bf \large $E_{vis}/\sqrt{s}$}
\put(-295,75){\rotatebox{90}{\bf \large $|\Sigma\bar{p}|/\sqrt{s}$}}
\caption{ The scatter plot of $E_{vis}/\sqrt{s}$ versus $|\Sigma\bar{p}|/\sqrt{s}$ for $J/\psi$ data. 
    The signal region, defined by 0.93 $\leq$ $E_{vis}/\sqrt{s}$ $\leq$ 1.10 and $|\Sigma\bar{p}|/\sqrt{s}$ $\leq$ 0.1, is shown as a box.}
\label{fig::openbox}
\end{center}
\end{figure}

Finally, 4 candidate events in the signal region are obtained from 225 million $J/\psi$ meson decays, which are shown in 
figure \ref{fig::openbox}. The detection efficiency for signal is determined to be (18.99 ± 0.12)\%. Based on a full simulated
$J/\psi$ MC sample whose size is 4 times of our experimental data, we find nineteen background events surviving in the signal 
region. This yields a predicted background of $N^{exp} = (4.75 \pm 1.09)$.

Considering the fact that only four events survived in signal region which is consistent with the number of potential backgrounds,
the pure signal events should be quite poor. Therefore, we set the upper limit on the branching fraction of $J/\psi \to e\mu$ 
based on the Feldman-Cousins method in which systematic uncertainties have been incorporated. The upper limit on the number of 
observed signal events at 90\% C.L. is obtained to be 6.15 by the POLE program \cite{ref::pole} inputing the number of expected 
background events, the number of observed events and the systematic uncertainty (5.84\%). The upper limit on the branching fraction 
is determined to be $\mathcal{B}(J/\psi \to e\mu)<1.6\times10^{-7}$.

\section{Prospects}
Here we make a full simlation to estimate the prospects of searching for cLFV signals in $J/\psi\to e\tau$ and $J/\psi\to \mu\tau$ 
based on the 1300 million $J/\psi$ sample.
The tracks in the final states of $J/\psi\to e\tau$ and $J/\psi\to \mu\tau$ are the same despite the momentum distribution, both 
have two opposite charge tracks and two missing tracks, so the analysis procedure of the two decays are similar. 
By analyzing the genreic MC sample of $J/\psi$ decay, we can found most of the backgrounds for the two decay modes are from 
$J/\psi\to\pi^+K_LK^-$, $J/\psi\to K_LK_L$, and $J/\psi\to K^{\star0}K^0$. After background supression, the detection efficiency 
is estimated to be 14\% and 19\% for $J/\psi\to e\tau$ and $J/\psi\to \mu\tau$, respectively. 
The sensitivities of the branching fraction are obtained to be $\mathcal{B}^{sensitivity}_{J/\psi\to e\tau}<6.3\times10^{-8}$ 
and $\mathcal{B}^{sensitivity}_{J/\psi\to\mu\tau}<7.3\times10^{-8}$ at 90\% C.L. with similar calculation method used in $J/\psi \to e\mu$. 

\section{Summary}
In summary, by analyzing 255 million $J/\psi$ data collected at the BESIII detector at the BEPCII collider, the charged Lepton 
Flavor Violation process is searched. Four signal events are observed which are consistent with the background estimation. 
As a result, we got the best upper limit in the world on the $J/\psi \to e\mu$ branching fraction at a 90\% CL. 
To get the prospects based on 1300 million $J/\psi$ data which has been accumulated by the BESIII experiment, we make a full 
MC simulation. The sensitivities on searching for cLFV signals in the $J/\psi\to e\tau$ and $J/\psi\to \mu\tau$ decays are 
estimated to be $6.3\times10^{-8}$ and $7.3\times10^{-8}$ at 90\% C.L., respectively, which will be the world best constraints.

\section*{Acknowledgments}
I would like to thank the committee of NuFact 2016 for the invitation and their excellent organizing.
This work is supported by the National Natural Science Foundation of China (NSFC) under Contracts No. 11475090 and No. 11005061.

\end{document}